\documentstyle[12pt]{article}
\renewcommand{\baselinestretch}{1.25}
\def\p0031{PSR B0031$-$07}
\def\xi{x_i}
\def\yi{y_i}
\def\xb{\left < x \right >}
\def\yb{\left < y \right >}
\def\sx{\sigma_x}
\def\sy{\sigma_y}
\def\delxi{\delta x_i}
\def\delyi{\delta y_i}
\def\delrxi{\frac{\delta \rho}{\delxi}}
\def\delryi{\frac{\delta \rho}{\delyi}}
\begin{document}
\large
\ 
\vskip 1.0in
\begin{tabbing}

\hspace*{3cm}\=sample column \=\hspace{15cm}\= \kill

{\bf Title}:\> Competing Drifting Radio Subpulses in PSR B0031$-$07 \\

\ \> \\

{\bf Authors}: \>  M. Vivekanand and B. C. Joshi. \\

\ \> \\

{\bf Address}: \>  National Centre for Radio Astrophysics \\
{\ } \> Tata Institute of Fundamental Research \\
{\ } \> Pune University Campus, Post Bag 3, Ganeshkhind \\
{\ } \> Pune 411007, INDIA. \\

\ \> \\

{\bf e-mail}: \> vivek@ncra.tifr.res.in, bcj@ncra.tifr.res.in \\

\end{tabbing}
\vfill
\eject
\normalsize
\sf
\title{Competing Drifting Radio Subpulses in PSR B0031$-$07}

\author{M. Vivekanand and B. C. Joshi}

\date{}

\maketitle
\begin{abstract}

\setlength{\textheight}{23.0cm}
\setlength{\textwidth}{13.5cm}
\setlength{\hoffset}{0.0cm}
\setlength{\voffset}{-3.0cm}
\renewcommand{\baselinestretch}{1.25}

\normalsize
\sf

The pair of drifting subpulses of the radio pulsar \p0031 appear
to be well separated from each other on the average, overlapping 
at less than one thirtieth of the peak energy. Strictly, the 
integrated profile of the pair (after removing the drift) does 
not show more than two subpulses; however this result is likely 
to be modified with the availability of more data. On the other 
hand, rare instances of three simultaneous subpulses in a single 
period have been noticed. This is consistent with the Ruderman 
and Sutherland model of the drifting phenomenon in which several 
sparks, uniformly spaced on a circle, rotate around the polar cap. 
The third subpulse appears to show up strongly when subpulses 
belonging to the primary drift bands weaken. Although this 
conclusion is based mainly on two reliable examples, it suggests 
an anti correlation of the subpulse energies. The normalised 
correlation coefficient between the energies of the two subpulses 
in a period is $-$0.16$\pm$0.02, after accounting for a known 
selection effect. Simulations suggest that this result could not 
be an artefact. This suggests the idea of competing drifting 
subpulses which, if true, will have important ramifications for 
the growth and breakdown of the particle acceleration zones in
pulsars.

\end{abstract}

\noindent {\Large \bf Subject Headings}: pulsar --- \p0031 
--- drifting --- competing subpulses --- acceleration zones.

\section{Introduction}

Among radio pulsars, \p0031 shows the second most systematic form
of drifting subpulses. It has three distinct rates of subpulse 
drifting (Huguenin, Taylor \& Troland 1970; Vivekanand \& Joshi 
1997), while the most systematic drifter PSR B0809+74 has one 
single drift rate (Cole 1970; Lyne \& Ashworth 1983). All other 
drifting pulsars either have not so uniform drifting (PSR 
B2016+28) or systematic drifting for only a fraction of the 
time (PSR B1919+21) (Taylor, Manchester \& Huguenin 1975;
Manchester \& Taylor 1977). Since PSR B0809+74 and \p0031 are 
weak pulsars, it has not been possible so far to study in 
requisite detail the pulse emission properties that are drift 
independent. For example, Michel (1991) states at the bottom
of page 68 of his book ``.. it suggests that the subpulse 
continues to exist outside of the viewing window! Thus in 
Ruderman and Sutherland (1975) we have the concept of a full 
series of unseen subpulses marching around the polar caps. It 
would be interesting if this view could be tested somehow''.

This has been attempted for \p0031 in this article. The data 
consist of about 33,000 periods of high quality, obtained at 
327 MHz using the Ooty Radio Telescope (ORT), which has only a 
single polarisation. Several drift dependent properties of this 
pulsar have already been studied; for these details as well as 
for details of data acquisition, pre-processing, calibration, 
identifying the drift bands, etc, see Vivekanand \& Joshi (1997) 
and Vivekanand (1995). 

The method of this article is to remove the drift rate from the 
motion of the subpulses, and to study them as if they came from a 
non-drifting pulsar (see Figure 1). This is possible due to the 
high collecting area of ORT, so that individual drift bands 
can be identified and their slopes estimated in most of the 33,066 
periods (Vivekanand \& Joshi 1997). Only those pairs of adjacent 
drift bands were considered that have at least three common periods 
with recognizable subpulses in each band. Since the drift rate 
changes from band to band, a slope was fit to each drift band (see 
Vivekanand \& Joshi 1997 for details), and only those pairs were 
further retained in which the two slopes were consistent with 
each other (difference of the slopes should be less than the rms 
error on it). This reduced the usable data to about a 
tenth of the original, similar to section 5.2 of Vivekanand \& Joshi 
(1997). Then the data of each period were shifted so that the mid 
points between the two slopes were aligned at each period. The 
shifting was done by fourier transforming the data (64 point FFT, 
zero padding the non-data samples), multiplying by the requisite 
phase gradient, and then inverse transforming. The data in the 
on-pulse window was corrected for variations due to inter stellar 
scintillations in the standard manner. The data of each period was 
further normalised so that the off-pulse window rms was 1.0.

\section{The Drift-Removed Integrated Profile}

Figure 2 shows the integrated profile of 3891 periods each of which 
is shifted as mentioned in section 1. Of these 2411 periods were 
acquired using a sampling interval of 7.5 milli seconds (ms), and 
1480 periods were acquired using 5.5 ms. The latter data were 
resampled at 7.5 ms before integration; this was done by the 
standard technique of fourier transforming the data, expanding the 
spectrum by the factor 11, low-pass filtering, and finally 
compressing the spectrum by the factor 15 (5.5/7.5 = 11/15; see any 
standard textbook on signal processing such as Oppenheim \& Schafer 
1989). The phase 0.0 of the original and the resampled data were 
ensured to be identical (i.e., the mid point between the two slopes).

We ask two questions in Figure 2.

\subsection{Are The Two Subpulses Well Separated ?}

A parabola was fit to the seven data between phases $-$22.5 ms and
+22.5 ms in Figure 2. The minimum of the parabola is at phase 
$-$0.36$\pm$0.33 ms, which is consistent with the expectation. The 
ordinate of the parabola (in arbitrary units) at this phase is 
$-$0.49$\pm$0.24. Now, the mean value of the data outside the dashed 
lines in Figure 2 is $-$0.001, while its root mean square (rms) 
deviation is 0.018; the latter is consistent with the expected value 
of $1 / \sqrt{3891} = 0.016$, since the rms of the off-pulse window 
data was normalised to 1.0 for each period. The minimum of the 
parabola is at a negative value, and is technically 0.49/0.24 = 2.05 
standard deviations away from the mean value in the wings of Figure 2; 
so the two numbers are not very inconsistent. However the error 0.24 
is much larger than the rms value of 0.018 in the wings, so the above 
conclusion has to be read with caution. We can assume that the minimum 
value of the parabola is unlikely to be greater than $-$0.49 +
3.0 $\times$ 0.24 = 0.23, which is much less than one thirtieth of 
the peak values. We therefore conclude that the emission from the 
two subpulses is well separated, at least at the level of one 
thirtieth of the peak energy of the subpulses.

\subsection{Do We See A Third Subpulse in the Wings ?}

The integrated profile in Figure 2 within the dashed lines 
(phases between $-$60.0 ms and +60.0 ms; 17 samples) was fit
to a sum of two gaussians plus a constant, of the form

\begin{equation}
a \exp - \left (\frac{t - b}{c} \right )^2 + 
d \exp - \left (\frac{t - e}{f} \right )^2 + g.
\end{equation}
The best fit results are $a$ = 8.82$\pm$0.08, $b$ = $-$26.03$\pm$0.06 
ms, $c$ = 9.39$\pm$0.10 ms, $d$ = 8.05$\pm$0.07, $e$ = +24.95$\pm$0.07 
ms, $f$ = 10.44$\pm$0.12 ms, and $g$ = 0.02$\pm$0.03; the $\chi^2$ 
per degree of freedom is 1.6. The quantities of main interest here 
are the positions of the two peaks $b$ and $e$. The precision of 
these numbers was verified by also fitting a function of the form

\begin{equation}
\frac{A}{\left [ (t - B)^2 + C^2 \right ]^{4.6}} + 
\frac{D}{\left [ (t - E)^2 + F^2 \right ]^{4.6}} + G
\end{equation}
where the exponent 4.6 was found to give the best fit. This gives
nearly identical results for the fitted parameters.

The mid point between the two peaks lies at $0.5 \times (b + e)$ = 
$-$0.54$\pm$0.05 ms; this is consistent with the position of the 
minimum of the parabola of the previous section. Their separation 
is $(e - b)$ = 50.98$\pm$0.09 ms.

Now, Ruderman \& Sutherland (1975) suggest that the drifting 
subpulse phenomenon is due to equally spaced ``sparks'' drifting 
in the crossed electric and magnetic fields on the polar cap. If 
so, we expect to see subpulses at the phases $b$ $-$ 50.98 = 
$-$77.01$\pm$0.11 ms and $e$ + 50.98 = +75.93$\pm$0.11 ms (as 
suggested by Michel 1991); these subpulses are expected to be weak 
since they fall at the edges of the integrated profile. On the 
positive phase side the unseen subpulse is expected in the sample 
32; the power at this sample is 0.001$\pm$0.018. By including two 
adjacent samples on either side of this sample (i.e. a total of 3 
samples), the mean power becomes 0.017$\pm$0.010; by including four 
adjacent samples (i.e. a total of 5 samples) the mean power becomes 
0.005$\pm$0.008. Thus there is no significant enhancement of power 
in this region, as is evident from Figure 2.

On the negative side, the expected phase falls in sample 12. An
exercise similar to the one above gives the mean powers 0.027$\pm$0.018, 
0.007$\pm$0.010, and 0.023$\pm$0.008, respectively; however the 
last number includes sample number 14 which lies on the dashed line,
so it should be ignored.

Therefore one is forced to conclude that one does not see the third
or fourth drifting subpulses in the averaged data. 

However, this conclusion will most probably be modified with the 
availability of more data, which will reduce the rms in the wings. 
Indeed, this conclusion appears to be already invalid when one looks 
closely at some individual periods. Figure 3 shows one such example, 
period number 619. The figure shows four drift bands that commence 
and end at the period pairs (602, 611), (606, 619) (611, 624), and 
(618, 624), the last band continuing beyond period number 624. Periods 
613 and 619 clearly show three subpulses. In period 619 the subpulse 
at sample 32, belonging  to the principal drift band, has become weak, 
while the two adjoining subpulses have shown up stronger. The subpulses 
at samples 22 and 40 are the hitherto unseen subpulses; it is not
surprising that such subpulses are difficult to notice, since they 
occur at the edge of the integrated profile. 

This best illustrates the idea of competing subpulses. Due to the 
variable drift rate of successive drift bands, and due to pulse to pulse 
intensity fluctuations, it will be fruitless to do a more rigorous 
analysis on period 619. This data is not included in Figure 2 due to the 
stringent selection criterion. One can not think of any instrumental or 
similar cause for explaining the properties in Figure 3, particularly in 
period 619.

Periods 617 and 618 illustrate the extreme pulse-to-pulse intensity
variations suffered by \p0031. The former has a weak subpulse at sample
26.6, while the latter has no recognisable subpulse above the receiver
noise, and is most probably a null period.

Figure 4 shows another example of competing drifting subpulses, viz., 
period numbered 1314, which was also not included in Figure 2 due to the 
stringent selection criterion. The above two examples were found after a 
careful scrutiny of almost 30,000 periods of data. While only two such 
were found, there were a handful more of the kind exemplified by period 
613 in figure 3.

The next section explores the idea of competing subpulses in greater
detail.

\section{Anti Correlation of Subpulse Energies}

One method of studying the idea of competing subpulses is to look
for anti correlation between their energies. 

At first glance the energies of the pairs of drifting subpulses are 
indeed anti correlated, but on account of weighting due to their
position within the the integrated profile. A typical subpulse enters 
a drift band at the right edge of the integrated profile, where its 
energy is low. As it drifts towards the left, it moves towards the 
center of the integrated profile, so its energy increases. After it 
crosses the peak of the integrated profile and drifts further to the 
left, its energy begins to decrease. The same thing happens to the 
companion subpulse, but with a delay, since that belongs to an adjacent 
drift band. It is easy to see that while one of the subpulses gains in 
strength, the other weakens, and vice versa (see Figure 1). This leads 
to an anti correlation of the observed subpulse energies. This selection 
effect has to be accounted for before studying the correlation of 
subpulse energies.

First, a shifted integrated profile of the kind in Figure 2 was formed 
for each data file. After identifying the limiting samples for the two 
subpulses (i.e., between the phase 0.0 and the respective dashed lines
in Figure 2), their energies $e1$ and $e2$ in each period were obtained 
by integrating the powers in the appropriate samples, and multiplying 
by the sampling interval for that file. Before shifting the data of 
each period, the values of the original integrated profile (i.e., of the
un-shifted data) were also estimated at the mean positions of the two 
subpulses, using linear interpolation; the method of obtaining the mean 
positions of the subpulses is described in Vivekanand \& Joshi (1997). 
These are the so called weights $w1$ and $w2$ for the two subpulse 
energies, which are clearly highly anti correlated. Since intensity 
variations due to inter stellar scintillations are removed in the data, 
one obtains an integrated profile scaled to a fixed energy across the 
several data files. It is in the normalised subpulse energies $e1/w1$ and 
$e2/w2$ that one should look for the anti correlation.

Figure 5 shows a plot of $\log_{10}(e2/w2)$ against $\log_{10}(e1/w1)$ 
for 3867 periods; the logarithmic plot exagerrates the anti
correlation for better viewing. The correlation coefficient $\rho$ is
defined as

\begin{equation}
\rho = \left < 
\frac{ \left [ \frac{e1}{w1} - \left < \frac{e1}{w1} \right > \right ]
\left [ \frac{e2}{w2} - \left < \frac{e2}{w2} \right > \right ] }
{ \sigma_{1} \sigma_{2} }
\right >,
\end{equation}
where $\left < ... \right >$ is the mean value and $\sigma_1$ and 
$\sigma_2$ are the standard deviations of $\frac{e1}{w1}$ and 
$\frac{e2}{w2}$, respectively. It is $-$0.037$\pm$0.023 for the entire data 
of Figure 5 (appendix A derives the error on $\rho$). The dashed box in 
Figure 5 excludes 307 points (about 8\% of the data) which are in the 
extremes. It is chosen by the criterion that both ${ \left [ \frac{e1}{w1} 
- \left < \frac{e1}{w1} \right > \right ] }/ { \sigma_{1} }$ and ${ \left 
[ \frac{e2}{w2} - \left < \frac{e2}{w2} \right > \right ] }/ { \sigma_{2} }$ 
be less than 4.5 in absolute value, and that both $\log_{10}(e2/w2)$ and 
$\log_{10}(e1/w1)$ be greater than $-$3.75; the value of $\rho$ is not 
sensitive to small changes in these parameters. The $\rho$ for the data 
within the box is $-$0.163$\pm$0.023, which is significantly less than zero. 
Clearly the above anti correlation is not due to the extreme data in Figure 
5. In fact, $\rho$ for the data outside the box turns out to be $-$0.08$\pm$0.07, 
which means that the extreme data are actually reducing the absolute value of 
$\rho$. These numbers have been carefully verified. For example, by excluding
merely four extreme data, $\rho$ changes from $-$0.037 to $-$0.114, mainly because
$\sigma_1$ changes from 0.049 to 0.033 and $\sigma_2$ changes from 0.049
to 0.019.

Since the data is spread over more than three orders of magnitude, we
also obtained the Spearman's rank correlation coefficient, which turns
out to be $-$0.66 for the data in the box in Figure 5; the Kendall rank 
correlation coefficient is $-$0.51. The rank correlation of the data outside
the box is also high ($-$0.74 and $-$0.49, respectively), due to the two well 
separated distribution of points across several orders of magnitude. The 
data within the dashed box sampled at 5.5 ms gives $\rho_{5.5}$ = 
$-$0.231$\pm$0.033, while that sampled at 7.5 ms gives $\rho_{7.5}$ = 
$-$0.137$\pm$0.031. Thus the anti correlation $\rho$ = $-$0.163$\pm$0.023 
in the entire data is a genuine feature. 

An important source of possible error in this section is the strong
anti correlation of the weights themselves. If an error has been made
in estimating the weights, could a residual small anti correlation 
remain in the normalised subpulse energies? For example, an error in 
estimating the mean position of the subpulses could result in erroneous 
weights. This has been investigated in appendix B by simulations using 
the actual integrated profile of \p0031. The conclusions are that (a) 
random errors in positions of the subpulses will not result in the 
observed anti correlation of pulse energies, as is intuitively obvious, 
and (b) even by making a systematic error of one sample in estimating 
the RELATIVE subpulse positions, (which is highly unlikely) the $\rho$ 
expected is +0.03, and it continues to be positive for several samples 
of error.

We therefore conclude that the energies of the pairs of drifting subpulses
in \p0031 are indeed anti correlated.

\section{Discussion}

The main results of this article are:

\begin{enumerate}

\item The average visible drift pattern of \p0031 consists of 
two well separated subpulses whose energies are anti correlated
with each other.

\item Occasionaly the third and fourth subpulses of the 
neighboring drift bands are also visible. This is consistent 
with the Ruderman \& Sutherland (1975) model of several equally
spaced sparks rotating on the polar cap.

\item These neighboring subpulses are not seen in the average 
visible drift pattern, most likely due to insufficient data.

\item It appears that the neighboring subpulses show up when the 
subpulses of the main drift bands weaken.

\item The above suggests that the subpulses of \p0031 are 
competing with each other for energy.

\end{enumerate}

These results appear to strongly support the Ruderman \& Sutherland 
(1975) picture of several equally spaced electric discharges that 
drift on the polar cap. Unfortunately ``... the entire theoretical 
foundation of that standard model had collapsed'' (Michel 1991, in
the preface of his book), mainly due to the low work function of the 
iron ($^{56}$Fe) ions (Flowers et al 1977), in spite of later efforts 
(Cheng \& Ruderman 1980) to salvage the situation. Thus a genuine 
need exists to re-investigate the theoretical basis of this model.

The anti correlation between the subpulse energies is significant,
although it is low, considering the fact that de-correlating effects
exist, such as pulse-to-pulse intensity fluctuations. The concept
of competing subpulses has so far not been suggested, to the best
of our knowledge. This result may have important bearing on the 
mechanisms of discharge of particle acceleration zones in pulsars,
irrespective of whether the mechanism is driven mainly by inverse
compton process or curvature radiation (for example, see Zhang \&
and Qiao 1996).

\vfill
\eject

\centerline{\large \bf APPENDICES}

%-------------------------------------------------------------------------
\appendix

\section{Standard Deviation of $\rho$}

Let $x$ and $y$ be gaussian random variables. Then

\begin{equation}
\rho = \frac{1}{N} \sum_i \frac{(\xi - \xb)(\yi - \yb)}{\sx \sy},
\end{equation}
where

\begin{equation}
\begin{array}{ll}
\xb = \frac{1}{N} \sum_i \xi, \: & \sx^2 = \frac{1}{N} \sum_i (\xi - \xb)^2, \\
\yb = \frac{1}{N} \sum_i \yi, \: & \sy^2 = \frac{1}{N} \sum_i (\yi - \yb)^2.
\end{array}
\end{equation}

For small changes in $\xi$ and $\yi$, the change in $\rho$ will be
given by

\begin{equation}
\begin{array}{ll}
\delrxi & = \frac{1}{N} \frac{1}{\sx}
\left [ \frac{\yi - \yb}{\sy} - \rho \frac{\xi - \xb}{\sx} \right ] \\
\delryi & = \frac{1}{N} \frac{1}{\sy}
\left [ \frac{\xi - \xb}{\sx} - \rho \frac{\yi - \yb}{\sy} \right ]
\end{array}
\end{equation}

Now,

\begin{equation}
\begin{array}{rl}
d \rho & = \sum_i \left [ \delrxi \delxi + \delryi \delyi \right ] \\
\Rightarrow \left < d \rho ^2 \right > & = \sum_i \left [ \left < \left (
\delrxi \delxi \right ) ^2 \right > + \left < \left ( \delryi \delyi 
\right ) ^2 \right > + 2 \left < \delrxi \delryi \delxi \delyi \right >
\right ]
\end{array}
\end{equation}
since 

\begin{equation}
\left < \delxi \delta y_j \right > = 0;  \: 
\left < \delxi (x_j - \xb) \right > = 0; \:
\left < \delyi (y_j - \yb) \right > = 0; \: {\mathrm{etc. \: for}} \: i \neq j.
\end{equation}

By straightforward algebra, and using $\left < \delxi^2 \right > = \sx^2$, 
$\left < \delyi^2 \right > = \sy^2$, and $\left < \delxi \delyi \right > =
\rho \sx \sy$, one obtains:

\begin{equation}
\begin{array}{ll}
\sum_i \left < \left ( \delrxi \delxi \right ) ^2 \right > &=
\frac{1}{N} \left ( 1 - \rho^2 \right ) \\
\sum_i \left < \left ( \delryi \delyi \right ) ^2 \right > &=
\frac{1}{N} \left ( 1 - \rho^2 \right ) \\
\sum_i \left < \delrxi \delryi \delxi \delyi \right > & = 
- \frac{1}{N} \rho^2 \left ( 1 - \rho^2 \right ).
\end{array}
\end{equation}

Using these three terms one gets the standard deviation of the correlation 
coefficient:

\begin{equation}
\sigma_{\rho} = \sqrt{ \left < d \rho ^2 \right > } = \sqrt{\frac{2}{N}}
\left ( 1 - \rho^2 \right ).
\end{equation}

\section{Error in $\rho$ Due to Error in Weights}

In this section the effect of estimating wrong weights upon $\rho$ is 
investigated. Figure 6 shows the integrated profile of \p0031 obtained 
from one of the data files; it is expected to be the same for the rest 
of the data. The profile is asymmetric; it was fit by a gaussian plus a
quadratic of the form

\begin{equation}
f(t) = a \exp - \left (\frac{t - b}{c} \right )^2 + d + e t + f t^2
\end{equation}
separately to the left and right portions of the data. The best fit 
parameters for the left side are: $a$ = 4.00$\pm$0.01, $b$ = 0.00$\pm$0.02
ms, $c$ = 19.75$\pm$0.11 ms, $d$ = $-$0.27$\pm$0.05, the last two parameters 
being negligible; the $\chi^2$ per degree of freedom was 1.2. The values
for the right side are $a$ = 3.78$\pm$0.07, $b$ = $-$2.80$\pm$0.66 ms, 
$c$ = 23.21$\pm$0.39 ms, $d$ = 0.09$\pm$0.04; the $\chi^2$ per degree of 
freedom was 1.3. The most important difference is that the rms width on 
the right side is larger.

Next, random positions $t$ of the first subpulse were generated, uniformly 
distributed over a range $-T/2$ and $+T/2 - \tau$, where $\tau$
is the separation between the two subpulses (in the terminology of drifting
pulsars $\tau$ would be equivalent to P2). Although several values of $T$ 
and $\tau$ were tried, the results quoted in this section will refer to the
choice $T$ = 110 ms, and $\tau$ = 49.5 ms, which are the appropriate
numbers for \p0031. Then the two weights would be $w1 = f(t)$ and $w2 =
f(t + \tau)$. A third weight was also generated, viz., $w3 = f(t + \tau + 
\tau^\prime)$, where $\tau^\prime$ could be a random number, a fixed number, 
etc; $\tau^\prime$ would model the kind of error one has made in estimating
the weights. The idea is that the observed subpulse energies are the true 
energies $e1$ and $e2$ weighted by the correct values of the integrated profile 
$w1$ and $w2$; one would thus measure the energies $e1 \times w1$ and $e2 
\times w2$. However, one would correct for this effect using the weights $w1$ 
and $w3$, instead of $w2$; the fractional error in the pulse energies would be 
$w2/w3$. 

For each choice of $T$, $\tau$ and $\tau^\prime$, one thousand random numbers 
(uniformly distributed) were generated for $t$, $e1$ and $e2$. The weights $w1$, 
$w2$ and $w3$ were also computed using the model profile of Figure 6. Then
the normalised correlation coefficient was found between $e1$ and $e2 \times
w2 / w3$. For small values of $\tau^\prime$ the weights $w2$ and $w3$ are
highly positively correlated, irrespective of the sign of $\tau^\prime$; this 
is intuitively obvious since for $\tau^\prime$ approaching zero, the two weights 
must be the same.

The numerical code was checked using a Gaussian integrated profile $f(t) = 1 / 
\sqrt{\pi \sigma} \times \exp - \left ( {t}/{\sigma} \right )^2$, for which
one analytically obtains

\begin{equation}
\left < f(t) f(t + \tau) \right > = \left [ \sqrt{\frac{2}{\pi}} \frac{1}{(T - 
\tau)\sigma} \right ] \exp - \left (\frac{\tau}{2\sigma} \right )^2 
\mathrm{erf} \left ( \frac{T - \tau}{\sqrt{2}\sigma} \right )
\end{equation}
where $\mathrm{erf}(x)$ is the error function defined as $1/\sqrt{\pi} \int_0^x 
\exp-x^2$.

The results are: $\rho$ is negligible for random values of 
$\tau^\prime$; for a wide choice of the input parameters $\rho$ is about 
$\approx$ 0.02 for a fixed $\tau^\prime$ of 5.5 ms, which is one sample for this 
data file, and highly unlikely to occur for all periods. For $\tau^\prime$ = 55 
ms, $\rho$ increases to 0.06. The conclusion is that even a small systematic 
error in estimating the RELATIVE positions of the subpulses will not explain the 
observed anti correlation in section 3; if at all the error will cause a small
positive correlation.

%-------------------------------------------------------------------------

\vfill
\eject

\noindent {\bf References}

\begin{enumerate}

\item Cheng, A. F. \& Ruderman, M. A. 1980, ApJ, 235, 576

\item Cole, T. W. 1970, Nature, 227, 788

\item Flowers, E. G. et al 1977, ApJ, 215, 291

\item Huguenin, G. R., Taylor, J. H., \& Troland, T. H. 1970, ApJ, 
162, 727

\item Lyne, A. G. \& Ashworth, M. 1983,  MNRAS,  204, 519

\item Manchester, R. N. \& Taylor, J. H. 1977, Pulsars (San Fransisco,
W. H. Freeman \& Company)

\item Michel, F. C. 1991, Theory of Neutron Star Magnetospheres 
(Chicago, University of Chicago Press)

\item Oppenheim, A. V. \& Schafer, R. W. 1989, Discrete-Time Signal
Processing (New Jersy, Prentice-Hall)

\item Ruderman, M. A. \& Sutherland, P. G. 1975, ApJ, 196, 51

\item Taylor. J. H., Manchester, R. N. \& Huguenin, G. R. 1975,
ApJ, 195, 513

\item Vivekanand, M. 1995,  MNRAS, 274, 785

\item Vivekanand, M. \& Joshi, B. C. 1997,  ApJ, 477, 431

\item Zhang, B. \& Qiao, G. J. 1996, A\&A, 310, 135

\end{enumerate}

\vfill
\eject

\noindent {\bf Figure Captions}

\begin{enumerate}

\item A typical pair of adjacent drift bands of \p0031, observed
on 10 June 1995. The left frame shows the pairs of drifting
subpulses for five consecutive periods. The right frame shows 
their mean positions along with error bars, and the best fit
line (dashed) through each drift band (see Vivekanand and Joshi 
1997 for details). Data of each period is shifted so that the 
point midway between the dashed lines is aligned. The on-pulse 
window is from samples 23 to 41 (inclusive); the rest of the data 
belongs to the off-pulse window. The sampling interval is 7.5 milli 
seconds.

\item Top frame shows the integrated profile of the drift removed 
subpulses, aligned such that the mid point between the two drift 
bands falls at the phase 0.0. The bottom frame is an enlarged version 
of the former, focusing on the noise in the wings.

\item Data for twenty three periods of \p0031 obtained on 5 Dec
1993, using a sampling interval of 5.5 ms. The period numbered
619 best illustrates the concept of competing drifting subpulses.
The on-pulse window is from samples 18 to 42 (inclusive).

\item Data for fourteen periods of \p0031 obtained on 25 Oct 1992, 
using a sampling interval of 5.5 ms. The period numbered 1314 
behaves similar to period 619 of figure 3. The on-pulse window is 
from samples 19 to 44 (inclusive).

\item Plot of $\log_{10}(e2/w2)$ against $\log_{10}(e1/w1)$ for 
3867 out of the 3891 periods of Figure 2. The box with dashed 
lines excludes some extreme data.

\item The continuous curve is the integrated profile $f(t)$ of 
3360 periods of \p0031 observed on 25 Oct 1992. The left and 
right portions have been fit to separate gaussians; the best fit 
models are represented by the crosses and the stars, respectively.
The abscissa is time $t$ in ms. A and B are the positions of a
pair of drifting subpulses, separated by the time $\tau$ ms.

\end{enumerate}

\end{document}